\documentclass{rspublic}

\usepackage{amsmath}
\usepackage{graphicx}

\numberwithin{equation}{section}

\begin{document}

\title[Geometry of C\u{a}lug\u{a}reanu's theorem]{Geometry of C\u{a}lug\u{a}reanu's theorem}

\author[M. R. Dennis and J. H. Hannay]{M. R. Dennis$^1$ and J. H. Hannay$^2$}

\affiliation{$^1$ School of Mathematics, University of Southampton, Highfield, Southampton SO17 1BJ, UK \\
$^2$ H.H. Wills Physics Laboratory, University of Bristol, Tyndall Avenue, Bristol BS8 1TL, UK}

\label{firstpage}

\maketitle

\begin{abstract}{Ribbon, Linking number, Twist, Writhe, Gauss map, Curve framing}
A central result in the space geometry of closed twisted ribbons is C\u{a}lug\u{a}reanu's theorem (also known as White's formula, or the C\u{a}lug\u{a}reanu-White-Fuller theorem). 
This enables the integer linking number of the two edges of the ribbon to be written as the sum of the ribbon twist (the rate of rotation of the ribbon about its axis) and its writhe. 
We show that twice the twist is the average, over all projection directions, of the number of places where the ribbon appears edge-on (signed appropriately) - the `local' crossing number of the ribbon edges. 
This complements the common interpretation of writhe as the average number of signed self-crossings of the ribbon axis curve. 
Using the formalism we develop, we also construct a geometrically natural ribbon on any closed space curve - the `writhe framing' ribbon.
By definition, the twist of this ribbon compensates its writhe, so its linking number is always zero.
\end{abstract}

\section{Introduction}\label{sec:int}

A simple generalization of a closed space curve is the notion of a ribbon. 
An ideal narrow ribbon in three-dimensional space is specified by the position of one edge at each point along its length, together with the unit normal vector to the ribbon at each point of this edge.  
If the ribbon is a closed loop (with two faces, not one as a M{\"o}bius band), then the two edges are non-intersecting closed curves in space which may wind around each other if the looped ribbon is twisted.  
Such ideal twisted ribbon loops are important in applications, for instance modelling circular duplex DNA molecules (Fuller 1971, 1978; Pohl 1980; Bauer \textit{et al.} 1980; Hoffman \textit{et al.} 2003), magnetic field lines (Moffatt \& Ricca 1992), phase singularities (Winfree \& Strogatz 1983; Dennis 2004), rotating body frames (Hannay 1998; Starostin 2002), and various aspects of geometric phase theory (Chiao \& Wu 1986; Kimball \& Frisch 2004).

A fundamental result in the geometry of twisted, closed ribbon loops is C\u{a}lug\u{a}\-rea\-nu's theorem (C\u{a}lug\u{a}reanu 1959, 1961; Moffatt \& Ricca 1992) (also referred to as White's formula (White 1969; Pohl 1980; Kauffman 2001; Eggar 2000), and the C\u{a}lug\u{a}reanu-White-Fuller theorem (Adams 1994; Hoffman \textit{et al.} 2003), which is expressed as
\begin{equation}
   Lk = Tw + Wr.							\label{eq:cwf}
\end{equation}
$Lk$ is the topological linking number of the two edge curves; it is the classical Gauss linking number of topology (described, for example, by Epple (1998)). 
The theorem states that this topological invariant is the sum of two other terms whose proper definitions will be given later and which individually depend on geometry rather than topology: the twist $Tw$ is a measure of how much the ribbon is twisted about its own axis, and the writhe $Wr$ is a measure of non-planarity (and non-sphericity) of the axis curve.

The formula actually has a very simple interpretation in terms of `views' of the ribbon from different projection directions (this is hinted at in the discussion of Kauffman (2001)). 
Such interpretations of $Lk$ and $Wr$ go back to Fuller (1971, 1978) and Pohl (1968{\em a}, {\em b}), but do not seem to have been extended to $Tw.$ 
This is our first result.
Our second uses this picture to construct, for any curve, a particular ribbon on it which has zero $Lk.$ 
This is the writhe framing ribbon, anticipated algebraically by J. H. Maddocks (private communication, unpublished notes; also see Hoffman \textit{et al.} 2003).

\begin{figure}
\begin{center}
\includegraphics*[width=8cm]{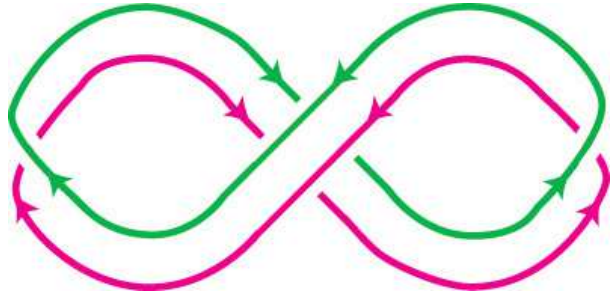}
\end{center}
    \caption{A projection of a twisted ribbon exhibiting the types of crossing described in the text. 
    The two edges of the ribbon are represented in different colours, and their linking number $Lk$ is +1. 
    At the left and right sides of the figure, when the ribbon is edge-on, there are `local crossings' (a `right-handed' or positive crossing on the right, a `left-handed' negative crossing on the left). 
    Two more positive crossings occur in the middle (where the ribbon crosses over itself); these correspond to the two crossings of the ribbon edges with themselves (i.e. a nonlocal crossing).}
    \label{fig:loop}
\end{figure}

The topological argument can be paraphrased very simply.
Consider the two edges of a ribbon loop, for example that in figure \ref{fig:loop}.  
Viewing a particular projection as in the picture, there are a number of places where one edge crosses the other. 
A positive direction around the ribbon is assigned arbitrarily, so that the two edges can be given arrows. 
At each crossing between the two edge curves, a sign ($\pm$) can be defined according to the the sense of rotation of the two arrows at the crossing ($+1$ for right handed, $-1$ for left handed, as shown in figure \ref{fig:loop}).
Summing the signs ($\pm 1$) of the crossings gives twice the linking number $Lk$ of the ribbon edges (the sign of $Lk$ is positive for a ribbon with planar axis and a right-handed twist, and is negative for a planar ribbon with a left-handed twist).
Since the ribbon is two-sided, the total number of crossings must be even, ensuring that $Lk$ is an integer.

The crossings between the two edge curves naturally fall into two types: `local,' which will be associated with $Tw,$ and `nonlocal,' which will be associated with $Wr.$  
Local crossings are where the ribbon is edge-on to the viewing direction: one edge of the ribbon is crossing its own other edge. 
If the ribbon is arbitrarily narrow, then the two edges are indefinitely close in space as they cross in the projection.  
Nonlocal type crossings between the two edges are caused by the ribbon crossing over itself.
More precisely, they occur when a single edge curve of the ribbon crosses over itself. 
This does not itself count towards the linking number, but such a crossing implies that there are two crossings with the other edge close by, as in the centre of figure \ref{fig:loop}.  
These two counts, the signed local crossings and signed nonlocal crossings (which come in pairs as described) add up to twice $Lk$ by definition. 
These quantities are integers, but depend on the choice of projection direction (although their sum does not).
It is well known (Fuller 1971, 1978; Adams 1994; Kauffman 2001) that the number of self-crossings of the ribbon axis curve, signed appropriately, and averaged democratically over the sphere of all projection directions, equals the writhe.
Therefore, the signed sum of nonlocal crossings between opposite edges is twice the writhe.

We claim that the local crossing counterpart (that is, the average over all projection directions of the signed local crossing number) is twice $Tw.$
The C\u{a}lug\u{a}reanu theorem then follows automatically.
This is described in the following section.
Section \ref{sec:formalism} is a review of standard material formalising the notions of linking and writhe; this is used in section \ref{sec:writhe} to construct the natural writhe framing of any closed curve.

The interpretation of twist and writhe in terms of local and nonlocal crossings gives insight into a common application of the theorem, namely `supercoiling' of elastic ribbons (such as DNA, or telephone cords; see, for example, Adams (1994), Pohl (1980), Bauer \textit{et al.} (1980) and Hoffman \textit{et al.} (2003)).
Repeated local crossings (i.e. a high twist) are energetically unfavourable, whereas nonlocal crossings (represented by writhe), where the ribbon repeatedly passes over itself (`supercoiling'), are preferred elastically.

\section{Local crossings, twist, and a proof of the C\u{a}lug\u{a}reanu theorem}\label{sec:proof}

In order to describe the argument more formally, it is necessary to introduce some notation.
We will represent the edges of the narrow ribbon by two closed curves, $\mathcal{A}$ and $\mathcal{B},$ whose points are $\mathbf{a}(s)$ and $\mathbf{b}(s)$ (and where no confusion will ensue, just $s$). 
$s$ is an arbitrary parameterization (giving the sense of direction along the curves), and $\rd \bullet/\rd s$ is denoted $\dot{\bullet}.$ 
$\mathcal{A}$ will be referred to as the {\em axis curve} and will play the primary role of the two curves. 
Its unit tangent $\mathbf{t}(s)$ is proportional to $\dot{\mathbf{a}}(s).$ 
The other curve $\mathcal{B}$ will be regarded as derived from $\mathcal{A}$ via a {\em framing} of $\mathcal{A},$ by associating at each point $\mathbf{a}(s)$ a unit vector $\mathbf{u}(s)$ perpendicular to the tangent ($\mathbf{t}\cdot\mathbf{u} = 0$). 
Then we define
\begin{equation}
\mathbf{b}(s) = \mathbf{a}(s) + \varepsilon \mathbf{u}(s)					\label{eq:bdef}
\end{equation}
for $\varepsilon$ arbitrarily small, ensuring that the ribbon nowhere intersects itself.
$Tw$ may now be defined formally, as the integral around the curve $\mathcal{A}$ of the rate of rotation of $\mathbf{u}$ about $\mathbf{t}:$ 
\begin{equation}
Tw = \frac{1}{2\pi} \int_{\mathcal{A}} \rd s \, (\mathbf{t} \times \mathbf{u}) \cdot \dot{\mathbf{u}}.			\label{eq:twdef}
\end{equation}
(We adopt the convention of writing the integrand at the end of an integration throughout.)
$Tw$ is local in the sense that it is an integral of quantities defined only by $s$ on the curve, and clearly depends on the choice of framing (ribbon).
A simple example of a framing is the {\em Frenet framing}, where $\mathbf{u}(s)$ is the direction of the normal vector $\dot{\mathbf{t}}(s)$ to the curve (where defined).
This framing plays no special role in the following.

As described in the introduction, the crossings in any projection direction can be determined to be either nonlocal (associated with writhe) or local.
Choosing an observation direction $\mathbf{o},$ there is a local crossing at $\mathbf{a}(s)$ (the ribbon appears edge-on), when $\mathbf{o}$ is linearly dependent on $\mathbf{t}(s)$ and $\mathbf{u}(s),$ i.e. there exists some $\theta$ between 0 and $\pi$ such that
\begin{equation}
   \mathbf{o} = \mathbf{t} \cos \theta + \mathbf{u} \sin \theta \qquad \hbox{at a local crossing.}
   \label{eq:local}
\end{equation}
We therefore define the vector
\begin{equation}
   \mathbf{v}(s, \theta) = \mathbf{t}(s) \cos \theta + \mathbf{u}(s) \sin \theta
   \label{eq:vdef}
\end{equation}
dependent on parameters $s,$ labelling a point of $\mathcal{A},$ and angle $\theta$ with $0 \le \theta \le \pi.$

We claim that twice $Tw$ is the average, over all projection directions $\mathbf{o},$ of the local crossing number.
For each $\mathbf{o},$ the local crossing number is defined as the number of coincidences of $\mathbf{v}$ with $\mathbf{o}$ or $-\mathbf{o}.$
$Tw$ itself (rather than its double) can therefore be determined by just counting coincidences of $\mathbf{v}$ with $\mathbf{o}$ (not $-\mathbf{o}$). 
The sign of the crossing is determined by which way the tangent plane of the
ribbon at $s$ sweeps across $\mathbf{o}$ as $s$ passes through the edge-on position (which side of the ribbon is visible before the crossing and which after).  
Thus the vector ($\partial_{\theta} \mathbf{v} \times \partial_s \mathbf{v}$) is either parallel or antiparallel to $\mathbf{v},$ and this decides the crossing sign: i.e. the sign of $(\partial_{\theta} \mathbf{v} \times \partial_s \mathbf{v})\cdot \mathbf{v}$ is minus the sign of the crossing.
This sign is opposite to the usual crossing number (described, for instance, in the next section), because $\theta$ increases in the opposite direction to $s$ along the ribbon.
The average over all projection directions $\mathbf{o}$ can be replaced by an integral over $s$ and $\theta,$ since the only projection directions which count are those for which there exist $s, \theta$ such that $\mathbf{o} = \mathbf{v}(s,\theta).$
This transformation of variables gives rise to a jacobian factor of $|(\partial_{\theta} \mathbf{v} \times \partial_s \mathbf{v})\cdot\mathbf{v}|$ (the modulus of the quantity whose sign gives the crossing sign). 
Our claim (justified in the following), that $Tw$ is the spherical average of crossing numbers, is therefore 
\begin{equation}
   Tw = \frac{1}{4\pi} \int_{\mathcal{A}} \rd s \int_0^{\pi} \rd \theta \,	(\partial_{\theta} \mathbf{v} \times \partial_s \mathbf{v})\cdot\mathbf{v}.
   \label{eq:twdef1}
\end{equation}

The two expressions for twist, equations (\ref{eq:twdef}) and (\ref{eq:twdef1}), are equal; this can be seen by integrating $\theta$ in equation (\ref{eq:twdef1}),
\begin{multline}
   \frac{1}{4\pi} \int_0^{\pi}  \rd \theta   \, (\partial_{\theta} \mathbf{v} \times \partial_s \mathbf{v})\cdot\mathbf{v} \\
   \begin{aligned}
   &= \frac{1}{4\pi} \int_0^{\pi} \rd \theta \,((\mathbf{t} \cos \theta + \mathbf{u} \sin \theta) \times (-\mathbf{t} \sin \theta + \mathbf{u} \cos \theta) ) \cdot (\dot{\mathbf{t}} \cos \theta + \dot{\mathbf{u}} \sin \theta)  \\
   &= \frac{1}{4\pi} \int_0^{\pi} \rd \theta \, (\cos^2 \theta + \sin^2 \theta) (\mathbf{t} \times \mathbf{u})\cdot (\dot{\mathbf{t}} \cos \theta + \dot{\mathbf{u}} \sin \theta) \\
   &= \frac{1}{2\pi} (\mathbf{t} \times \mathbf{u})\cdot\dot{\mathbf{u}},
   \end{aligned}
   \label{eq:twder}
\end{multline}
which is the integrand of equation (\ref{eq:twdef}).
Thus the two expressions for twist, the conventional one (equation (\ref{eq:twdef})) and the local crossing count averaged over viewing directions (equation (\ref{eq:twdef1})), are the same.
The preceding analysis of $Tw$ is reminiscent of the more abstract analysis of Pohl (1980), albeit with a different interpretation.

Using the notion of direction-averaged crossing numbers, the C\u{a}lug\u{a}reanu theorem follows immediately, as we now explain.
For a sufficiently narrow ribbon, it is straightforward to decompose the crossings between different edges into local and nonlocal for each projection of the ribbon.
It is well known (Fuller (1971, 1978), Adams (1994)) that writhe $Wr$ equals the sum of signed nonlocal crossings, averaged over direction, i.e. twice the average of self-crossings of the axis curve with itself (a proof is provided in the next section).
The average of local crossings (between different edges, counting with respect to both $\mathbf{o}$ and $-\mathbf{o}$), has been shown to equal twice the twist $Tw$ defined in equation (\ref{eq:twdef}).
The sum of local plus nonlocal crossings is independent of the choice of projection direction, and is twice the linking number $Lk$ of the two curves.
So, averaging the crossings over the direction sphere, $Lk = Tw + Wr.$

Any ambiguity as to whether a crossing is local or nonlocal arises only when the projection direction $\bf{o}$ coincides with the (positive or negative) tangent $\pm \mathbf{t}.$
However, the set of such projection directions is only one-dimensional (parameterised by $s$), and so does not contribute (has zero measure) to the total two-dimensional average over the direction sphere.

\section{Formalism for writhe and linking number}\label{sec:formalism}

We include the present section, which reviews known material (e.g. Fuller 1971, 1978; Adams 1994; Kauffman 2001; Hannay 1998), to provide some formal geometrical tools for the next section, as well as providing further insight into the proof from the previous section.

The linking number $Lk$ between the curves $\mathcal{A}$ and $\mathcal{B}$ is related to the system of (normalised) {\em cross chords}
\begin{equation}
   \mathbf{c}_{\mathcal{A}\mathcal{B}}(s,s') = \frac{\mathbf{a}(s) - \mathbf{b}(s')}{| \mathbf{a}(s) - \mathbf{b}(s') |}.
   \label{eq:cabdef}
\end{equation}
$Lk$ is represented mathematically using Gauss's formula:
\begin{eqnarray}
Lk & =& \frac{1}{4\pi} \int_{\mathcal{A}} \rd s \int_{\mathcal{B}} \rd s' 
\frac{( \dot{\mathbf{a}}(s) \times \dot{\mathbf{b}}(s') )\cdot(\mathbf{a}(s) - \mathbf{b}(s'))}{| \mathbf{a}(s) - \mathbf{b}(s') |^3}  \nonumber \\
& =& \frac{1}{4\pi} \int_{\mathcal{A}} \rd s \int_{\mathcal{B}} \rd s' \,  (\partial_s \mathbf{c}_{\mathcal{A}\mathcal{B}} \times \partial_{s'} \mathbf{c}_{\mathcal{A}\mathcal{B}})\cdot \mathbf{c}_{\mathcal{A}\mathcal{B}}.		\label{eq:lk}
\end{eqnarray}
This bears some similarity to the formula for twist in equation (\ref{eq:twdef}).
$Lk$ is invariant with respect to reparameterization of $s$ and $s',$ and, of course, any topological deformation avoiding intersections. 
The domain of integration in equation (\ref{eq:lk}), $\mathcal{A} \times \mathcal{B},$ is the cross chord manifold (secant manifold) of pairs of points on the two curves, topologically equivalent to the torus. 
The mapping $\mathbf{c}_{\mathcal{A}\mathcal{B}}(s,s')$ takes this torus smoothly to the sphere of directions, with the torus `wrapping around' the sphere an integer number of times (the integer arises since the cross chord manifold has no boundary, and the mapping is smooth); the wrapping is a two-dimensional generalization of the familiar `winding number' of a circle around a circle.
This wrapping integer, the integral in equation (\ref{eq:lk}), is the linking number of the two curves (it is the degree of the mapping; see, for example, Madsen \& Tornhave (1997), Epple (1998)).

It is easy to see that this interpretation of $Lk$ agrees with that defined earlier in terms of crossings. 
Choosing the observation direction $\mathbf{o},$ the crossings in the projection are precisely those chords $\mathbf{c}_{\mathcal{AB}}(s,s')$ coinciding with $\mathbf{o}$ where the sign of the scalar triple product $(\partial_s \mathbf{c}_{\mathcal{A}\mathcal{B}} \times \partial_{s'} \mathbf{c}_{\mathcal{A}\mathcal{B}})\cdot \mathbf{c}_{\mathcal{A}\mathcal{B}}$ gives the sign of the crossing.
As with $Tw,$ a crossing is only counted in the integral when $\mathbf{c}_{\mathcal{AB}}(s,s')$ is parallel to $\mathbf{o}$ (not antiparallel).
Since the cross chord manifold is closed (has no boundary), the total sum of signed crossings does not depend on the choice of $\mathbf{o}.$

\begin{figure}
\begin{center}
\includegraphics*[width=11cm]{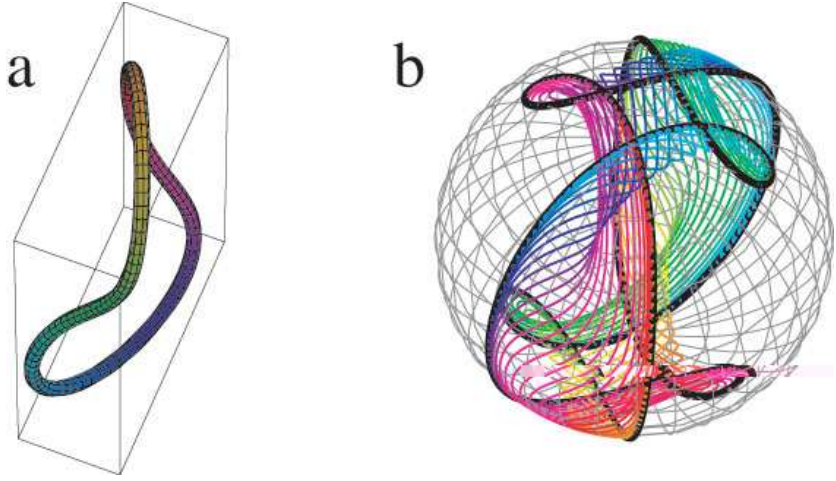}
\end{center}
    \caption{Illustrating the writhe mesh construction. 
    a) The curve on the torus $(-0.3 \cos(2s), \cos(s)(1+0.3 \sin(2s)), \sin(s)(1+0.3 \sin(2s)).$ 
    The points on the curve are represented by different colours on the colour wheel. 
    b) The writhe mesh for the curve in a). 
    The black lines are the tangent curves $\pm \mathbf{t}(s),$ and the coloured lines are the chord fans $\mathcal{C}_s,$ whose colours correspond to the points $s$ on the curve in a). 
    Note that the total writhe in this example is less than $4\pi$ (i.e. the writhe mesh does not completely cover the direction sphere).}
    \label{fig:writhe}
\end{figure}

Writhe has a similar interpretation to link. 
We define the chords between points of the same curve $\mathcal{A},$
\begin{equation}
   \mathbf{c}_{\mathcal{A}}(s,s') = \frac{\mathbf{a}(s) - \mathbf{a}(s')}{| \mathbf{a}(s) - \mathbf{a}(s') |},
   \label{eq:cadef}
\end{equation}
noting that, as $s'\to s$ from above, $\mathbf{c}_{\mathcal{A}} \to \mathbf{t}(s),$ and as $s'\to s$ from below, $\mathbf{c}_{\mathcal{A}} \to -\mathbf{t}(s).$
The writhe is the total area on the direction sphere traversed by the vector as $s$ and $s'$ are varied; this two-dimensional surface embedded on the sphere will be referred to as the {\em writhe mesh}:
\begin{eqnarray} 
Wr &=& \frac{1}{4\pi} \int_{\mathcal{A}} \rd s \int_{\mathcal{A}} \rd s'  
\frac{( \dot{\mathbf{a}}(s) \times \dot{\mathbf{a}}(s') )\cdot(\mathbf{a}(s) - \mathbf{a}(s'))}{| \mathbf{a}(s) - \mathbf{a}(s') |^3} \nonumber \\
&=& \frac{1}{4\pi} \int_{\mathcal{A}} \rd s \int_{\mathcal{A}} \rd s' \, (\partial_s \mathbf{c}_{\mathcal{A}} \times \partial_{s'} \mathbf{c}_{\mathcal{A}})\cdot \mathbf{c}_{\mathcal{A}}. \label{eq:wr}
\end{eqnarray}
The interpretation of crossings applies to the integral for writhe as well as link and twist; however, for writhe, the visual crossings are uniquely associated with a chord (the chord from $s$ to $s',$ and its reverse, each have separate associated viewing directions).
Thus, for writhe, each crossing is counted exactly once, and the average over projections is not divided by 2 (unlike link and twist).
(However, as shown in figure \ref{fig:loop}, at every nonlocal crossing of curve $\mathcal{A}$ with itself, there are two crossings (of the same sign) with the other curve.)

Unlike the cross chord manifold $\mathbf{c}_\mathcal{AB},$ the writhe mesh $\mathbf{c}_{\mathcal{A}}$ has a boundary: it is topologically equivalent to an annulus (i.e. a disk with a hole).
The two boundary circles map to the tangent indicatrix curves (i.e. the loops $\pm \mathbf{t}(s)$ on the direction sphere, as $s$ varies).
For each $s,$ the following locus on the sphere, referred to as the {\em chord fan}
\begin{equation}
   \mathcal{C}_s = \mathbf{c}_{\mathcal{A}}(s,s') \quad \hbox{(varying $s'$ from $s$ round to $s$ again)},
   \label{eq:csetdef}
\end{equation}
follows the directions of all the chords to the point $s,$ starting in the positive tangent direction $+\mathbf{t}(s),$ and ending at its antipodal point $-\mathbf{t}(s).$
Topologically, it is a `radial' line joining the two edges of the writhe mesh annulus, although on the direction sphere it may have self-intersections. 
In addition to the partial covering bounded by the tangent indicatrix curves $\pm\mathbf{t}(s),$ the total writhe mesh may cover the direction sphere an integer number of times.
The signed number of crossings, as a function of viewing direction $\mathbf{o},$ changes (by $\pm2$) when $\mathbf{o}$ crosses $\pm \mathbf{t}(s).$
The writhe integral (equation (\ref{eq:wr})) is its average value over all viewing directions.
The writhe mesh construction is illustrated in figure \ref{fig:writhe}.

The approach of section \ref{sec:proof} may also naturally be interpreted topologically on the direction sphere.
The vector $\mathbf{v}(s,\theta),$ dependent on two parameters, also defines a mesh on the direction sphere, the {\em twist mesh}.
For fixed $s,$ the locus of points on the twist mesh is a semicircle, 
\begin{equation}
   \mathcal{S}_s = \mathbf{v}(s,\theta) \qquad \hbox{($0 \le \theta \le \pi$)}
   \label{eq:ssetdef}
\end{equation}
whose endpoints are at $\pm\mathbf{t}(s)$ and whose midpoint is the framing vector $\mathbf{u}(s).$
As $s$ varies around the curve, the semicircle $\mathcal{S}_s$ sweeps out the solid angle $Tw.$
Like the writhe mesh, the twist mesh is topologically an annulus, with $\mathcal{S}_s$ the `radial' line labelled by $s;$ its boundary is again the set of tangent directions $\pm \mathbf{t}(s).$
Therefore, a topological visualization of $Tw + Wr$ is the union of the meshes for twist and writhe; each mesh has the same boundary, so the closure of the union is the join of two annuli along their boundary, topologically a torus.
This torus therefore wraps around the sphere the same number of times (has the same degree) as the cross chord manifold $\mathbf{c}_{\mathcal{AB}}.$

It is therefore possible to interpret the loop $\mathcal{L}_s$ on the manifold of cross chords $\mathcal{A}\times\mathcal{B},$ labelled by $s$ on $\mathcal{A}$ (varying $s'$ on $\mathcal{B}$), in terms of the twist semicircle $\mathcal{S}_s$ and the chord fan $\mathcal{C}_s.$ 
The loop $\mathcal{L}_s,$ mapped to the direction sphere, is the set of cross chords directions between fixed $\mathbf{a}(s)$ and all $\mathbf{b}(s')$ on $\mathcal{B}.$
When the ribbon is vanishingly thin, for $s'$ outside the neighbourhood of $s,$ the chords $\mathbf{c}_{\mathcal{AB}}(s,s')$ can be approximated by the chords of $\mathbf{c}_{\mathcal{A}}(s,s').$
When $s'$ is in the neighbourhood of $s,$ $\mathbf{c}_{\mathcal{AB}}(s,s')$ is approximated by $\mathbf{v}(s,\theta)$ for some $\theta$ (exact at $\theta = \pi/2$ when $s' = s$).
As $\varepsilon \to 0,$ these approximations improve, and $\mathcal{L}_s$ approaches the union of $\mathcal{C}_s$ and $\mathcal{S}_s.$
Link is therefore the area on the direction sphere swept out by the loop union of $\mathcal{C}_s$ and $\mathcal{S}_s$ for $s$ varying around the curve; since this family of loops generates the closed torus, the direction sphere is enveloped an integer number of times.
   
Thus, C\u{a}lug\u{a}reanu's theorem may be interpreted as a natural decomposition of the integrand in Gauss's formula (\ref{eq:lk}) in terms of the writhe mesh and twist mesh. 
This is very close, in a different language, to White's proof (White 1969, also see Pohl 1968{\em a}, and particularly Pohl 1980), where objects analogous to the writhe mesh and the twist mesh appears as boundaries to a suitably regularised (blown-up) 3-manifold of chords from $\mathcal{A}$ to points on surface of the ribbon with boundary $\mathcal{A}, \mathcal{B}.$

\section{The writhe framing}\label{sec:writhe}

In this section, we use the description of $Tw$ and $Wr$ to define, for any non-self-intersecting closed curve in space, a natural framing (i.e. ribbon) whose linking number is zero.
Such a framing is useful since it can be used as a reference to determine the linking number for any other framing: if $\mathbf{u}_0(s)$ represents this zero framing, and $\mathbf{u}(s)$ any other framing with linking number $Lk,$ then
\begin{equation}
   Lk = \frac{1}{2\pi} \int_{\mathcal{A}} \rd s\, \arccos \mathbf{u}\cdot\mathbf{u}_0.
   \label{eq:zerolinkref}
\end{equation}
Clearly the twist of a zero framing is equal to minus the writhe by C\u{a}lug\u{a}reanu's theorem, and therefore we will refer to our natural framing as the {\em writhe framing}.
Expressing $Lk$ in terms of an integral involving the difference between two framings is reminiscent of C\u{a}lug\u{a}reanu's original proof (C\u{a}lug\u{a}reanu 1959, 1961; also see Moffatt \& Ricca 1992), in which the twist is defined as the total number of turns the framing vector makes with respect to the Frenet framing around the curve (provided the normal to the curve is everywhere defined).
The total twist is therefore the sum of this with the integrated torsion around the curve (i.e. $Tw$ of the Frenet framing).
As stated before, the Frenet framing plays no role in our construction.

It is easy to construct an unnatural zero framing by taking any framing, cutting anywhere, and rejoining with a compensating number of twists locally. 
In contrast, the framing we construct here is natural in the canonical sense that, at any point $s,$ the definition of $\mathbf{u}_0(s)$ and its corresponding semicircle $\mathcal{S}_s$ (defined in equation (\ref{eq:ssetdef})) depends only on the view from $s$ of the rest of the closed curve, that is on the chord fan $\mathcal{C}_s$ (defined in equation (\ref{eq:csetdef})).
The rate of rotation of the resulting $\mathbf{u}_0$ exactly compensates the corresponding writhe integrand, that is, the integrands (with respect to $s$) of equation (\ref{eq:wr}) and the twist of the writhe framing (equation (\ref{eq:twdef})) are equal and opposite.

\begin{figure}
\begin{center}
\includegraphics*[width=12.15cm]{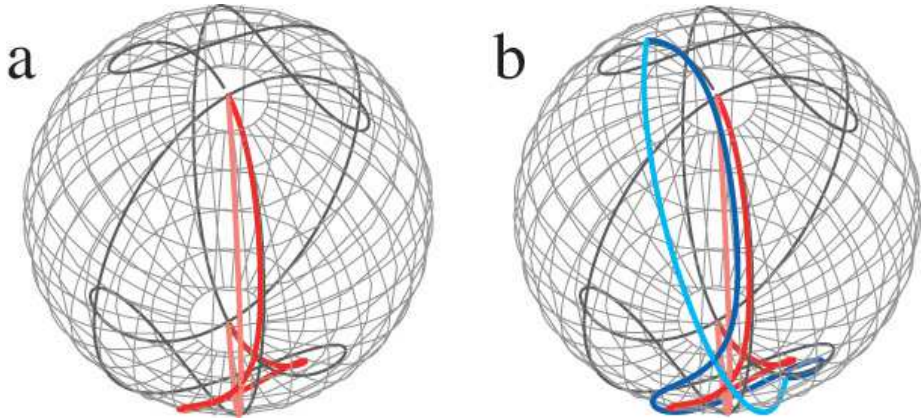}
\end{center}
    \caption{Illustrating the writhe framing construction, using the same example curve as figure \ref{fig:writhe}. 
    a) The chord fan for $s = 0,$ is represented in red, the semicicle which bisects its area represented in pink. 
    The tangent curves are also represented. 
    b) The same as part a), and in addition the chord fan with $s = -0.4$ is shown in dark blue, and its bisecting semicircle in light blue. 
    No area is swept out by these closed curves as $s$ evolves.}
    \label{fig:framing}
\end{figure}

This may be interpreted topologically as follows.
$\mathcal{C}_s$ and $\mathcal{S}_s$ correspond topologically to `radial' lines of their respective annular meshes; their union is a closed loop on the direction sphere.
The sum of the twist and writhe integrands, by the discussion in section \ref{sec:proof}, is the area swept out by this changing closed loop as $s$ develops; it was proved that the total area swept out by this loop is $4\pi Lk.$
The writhe framing construction of $\mathbf{u}_0$ below arranges that the loop has a constant area (say zero), and therefore the rate of area swept by this loop is zero as $s$ evolves, giving zero total area swept (which is $4\pi Lk$).

The direction of $\mathbf{u}_0(s)$ for the writhe framing is that for which the semicircle $\mathcal{S}_s$ bisects the chord fan $\mathcal{C}_s$ in the following sense. 
Since $\mathcal{C}_s$ is a curve on the direction sphere with endpoints $\pm\mathbf{t}(s),$ it may be closed by an arbitrary semicircle with the same endpoints (i.e. any rotation of $\mathcal{S}_s$ about $\pm\mathbf{t}(s)$).
The total closed curve encloses some area on the sphere (mod 4$\pi$); the (unique) semicircle which gives zero area (mod $4\pi$) defines $\mathbf{u}_0.$
Since the spherical area enclosed by this closed curve is contant, the curve does not sweep out any area as it evolves (since as much leaves as enters).
An example of the writhe framing is represented in figure \ref{fig:framing}.

The rate of area swept out by this curve is the integrand with respect to $s$  in the $Lk$ expression (equation (\ref{eq:lk})), (i.e. the sum of the $Tw, Wr$ integrands).
Since it has been shown that this is zero, the writhe framing $\mathbf{u}_0$ indeed has zero linking number.
Of course, although the choice of zero area is most natural, the writhe faming vector $\mathbf{u}_0$ could be defined such that the area between the chord fan $\mathcal{C}_s$ and the twist semicircle $\mathcal{S}_s$ is any fixed value - the important feature in the construction is that the area does not change with $s.$

\begin{acknowledgements}

We are grateful to John Maddocks for originally pointing out to us the problem of the writhe framing, and useful correspondence. MRD acknowledges support from the Leverhulme Trust and the Royal Society of London.

\end{acknowledgements}

\label{lastpage}

\end{document}